\documentclass[%
 aip,
 amsmath,amssymb,
 reprint,%
]{revtex4-1}

\usepackage{graphicx}
\usepackage{dcolumn}
\usepackage{bm}

\usepackage[utf8]{inputenc}
\usepackage[T1]{fontenc}
\usepackage{mathptmx}
\usepackage{etoolbox}

\makeatletter
\def\@email#1#2{%
 \endgroup
 \patchcmd{\titleblock@produce}
  {\frontmatter@RRAPformat}
  {\frontmatter@RRAPformat{\produce@RRAP{*#1\href{mailto:#2}{#2}}}\frontmatter@RRAPformat}
  {}{}
}%
\makeatother
\begin{document}

\preprint{AIP/123-QED}

\title[]{An FPGA-based Timing and Control System for the Dynamic Compression Sector}
\author{Shefali Saxena}
\email{shefali.saxena2103@gmail.com }%
\author{Daniel R. Paskvan}%
\author{Nicholas R. Weir}%
\affiliation{ 
Advanced Photon Source, Argonne National Laboratory, Lemont, IL 40639, USA
}%

\author{Nicholas Sinclair}
 \affiliation{%
Dynamic Compression Sector, Institute for Shock Physics,Washington State University, Argonne, Illinois 60439, USA
}%

\date{\today}

\begin{abstract}
A field programmable gate array (FPGA) based timing and trigger control system has been developed for the Dynamic Compression Sector (DCS) user facility located at the Advanced Photon Source (APS) at Argonne National Laboratory. The DCS is a first-of-its-kind capability dedicated to dynamic compression science. All components of the DCS laser shock station—x-ray choppers, single-shot shutter, internal laser triggers, and shot diagnostics—must be synchronized with respect to the arrival of x-rays in the hutch. A field-programmable gate array (FPGA) synchronized to the APS storage ring radio frequency (RF) clock (352 MHz) generates trigger signals for each stage of the laser and x-ray shutter system with low jitter. The system is composed of a Zynq FPGA, a debug card, line drivers and power supply. The delay and offsets of trigger signals can be adjusted using a user-friendly graphical user interface (GUI) with high precision. The details of the system architecture, timing requirements, firmware, and software implementation along with the performance evaluation are presented in this paper. The system offers low timing jitter (15.5 \emph{ps} r.m.s.) with respect to APS 352 MHz clock , suitable for the 50 \emph{ps} r.m.s. x-ray bunch duration at the APS.
\end{abstract}

\maketitle

\section{\label{sec1}Introduction}

The Dynamic Compression Sector is a user facility for synchrotron-based dynamic compression science studies.  The dynamic compression facility—consisting of gun-based impact facilities and a laser shock facility—is coupled to a high-flux x-ray beam and focuses on time-resolved x-ray diffraction, scattering and imaging measurements in materials subjected to dynamic compression and deformation. The DCS laser shock station allows users to compress millimeter-scale targets to multi-Mbar pressures over a duration of nanoseconds using the 100J UV laser system. During compression, x-ray diffraction [\onlinecite{Wang}] or EXAFS [\onlinecite{Das}] measurements from a single x-ray pulse provide atomic scale structural information, while VISAR (Velocity Interferometer System for Any Reflector ) interferometry provides velocity measurements on the continuum scale. The timing of the x-ray probe is critical to the success of an experiment, as the probe timing must correspond to a desired shock state in the sample (e.g. just as the shock reaches the back of the sample). A low-jitter triggering system for the synchronization of the DCS 100J laser to x-ray arrival at the sample allows users to precisely specify the timing of the x-ray probe.

 The Advanced Photon Source at the U.S. Department of Energy’s Argonne National Laboratory provides ultra-bright, high-energy storage ring-generated x-ray beams for research. The APS is a storage ring running with an internal clock of 352 MHz. The storage ring is operated in three common modes: 24 bunch mode, hybrid mode and 324 bunch mode. The details of each mode is provided in Table~\ref{tab:table1}. 

\begin{table*}
\caption{\label{tab:table1}The Advanced Photon Source (APS) storage ring operation modes, fill patterns, and bunch lengths.}
\begin{center}
\renewcommand\arraystretch{1.2}
\begin{tabular}{| >{\raggedright\arraybackslash}p{0.25\linewidth} | >{\raggedright\arraybackslash}p{0.6\linewidth} 
| >{\raggedright\arraybackslash}p{0.15\linewidth}|}
\hline
\textbf{Mode} & \textbf{Fill Pattern} & \textbf{Bunch Length (r.m.s.)}\\  
 \hline
24 bunch mode	
& 102 \emph{mA} in 24 singlets (single bunches) with a nominal current of 4.25 \emph{mA} and a spacing of 153 \emph{ns} between singlets.	
& 33.5 \emph{ps}
\\\hline
324 bunch mode	
& 102 \emph{mA} in 324 uniformly spaced singlets with a nominal single bunch current of 0.31 \emph{mA} and a spacing of 11.37 \emph{ns} between singlets.	
& 22 \emph{ps}
\\\hline
Hybrid mode	
& Total current is 102 \emph{mA}. A single bunch containing 16 \emph{mA} isolated from the remaining bunches by symmetrical 1.594 \emph{µs} gaps. The remaining current is distributed in 8 group of 7 consecutive bunches with a maximum of 11 \emph{mA} per group, a periodicity of 68 \emph{ns}, and a gap of 51 \emph{ns} between groups. The total length of the bunch train is 500 \emph{ns}.	
& Singlet: 50 \emph{ps} septuplets: 27 \emph{ps}
\\\hline
\end{tabular}
\end{center}
\end{table*}

The timing and control system for the DCS laser should be able to provide synchronization triggers and clocks with low jitter values in the range of a few tens of \emph{ps}. Meeting these timing specifications requires a carefully-considered design approach and state-of-the art hardware based on a field programmable gate array (FPGA). Several FPGA-based timing and trigger systems have been developed in the past for specific applications [\onlinecite{Akerib,Angelucci,Galli,Linden}]. The most important required characteristics are precise timing control and low jitter. Digital delay generators with low jitter have been extensively explored in the past [\onlinecite{Klep,Martin,Jang,Kalisz,Zhang}]. An FPGA-based timing and control system for the Large Hadron Collider (LHC) experiments has been described in [\onlinecite{Aloisio}], which demonstrates the results of phase noise and jitter analysis. Similarly, FPGA-based data acquisition systems provide significant advantages to nuclear physics experiments as they allow both reduction of dead-time \textemdash thus improving the throughput \textemdash and provide more accurate measurements [\onlinecite{Saxena, Ottanelli}]. 

In this work, an FPGA-based timing and control system is developed which is synchronized with the APS 352 MHz main RF clock and provides the trigger signals for all facets of the experiment. This includes all diagnostic measurement systems (diagnostic laser sources and oscilloscopes), x-ray shutter systems, and many components within the 100J laser system. The details of timing requirements, firmware-software implementation, system architecture, and performance evaluation are presented in the next sections.  

\section{\label{sec2}Timing and Control Requirements, Firmware and Software}

\subsection{\label{sec2a}Timing and Control Requirements}

Users of the DCS laser shock station may specify the timing of the x-ray probe pulse with respect to the arrival of the UV drive laser pulse on the sample. This delay is determined by the relative timing of the first stage of the laser and the arrival of an x-ray pulse on the sample. This timing, as well as the timing of all additional experimental components, must be synchronized to the APS RF clock. The jitter in the x-ray/laser synchronization should be much less than the transit time of a shockwave through a typical sample (1 to a few \emph{ns}), but since the x-ray pulse duration is 50 \emph{ps} r.m.s, it is not beneficial to engineer a solution with jitter much less than 10 \emph{ps}. 

Three x-ray shuttering systems are synchronized to isolate a single x-ray pulse per experiment, and the selection of this pulse sets the x-ray arrival time on target. The x-ray bunches are first chopped using a high heat load chopper (HHLC), which rotates at 82 Hz with an opening time of approximately 20 \emph{µs}. A single-shot shutter known as the milli-second (MS) shutter with a minimum 4 \emph{ms} window selects a single  sequence of 20 \emph{µs} bunches. The x-rays are then focused through a chopper (Forschungszentrum Jülich GmbH), called Jülich chopper, rotating at 987 Hz, which isolates a single x-ray pulse. 

In the DCS laser system, an electro-optical modulator, driven by an Arbitrary Waveform Generator (AWG), modulates a continuous wave fiber laser to generate seed laser pulses, which are then fed into further amplification stages. In the first stage of the laser, the FPGA must send a trigger must be sent at a frequency of 329 Hz to set the desired repetition rate of seed pulses into the first amplification stage. After the initial pulse slicing, several triggers running at 2.7 Hz with variable delay settings are generated to trigger the subsequent stages of the laser e.g., Multiple Frequency Modulation (Multi-FM), Regenerative amplifier, Pockels cells, as well as the triggers for the laser diagnostics. 

\begin{figure*}
\includegraphics[scale=0.5]{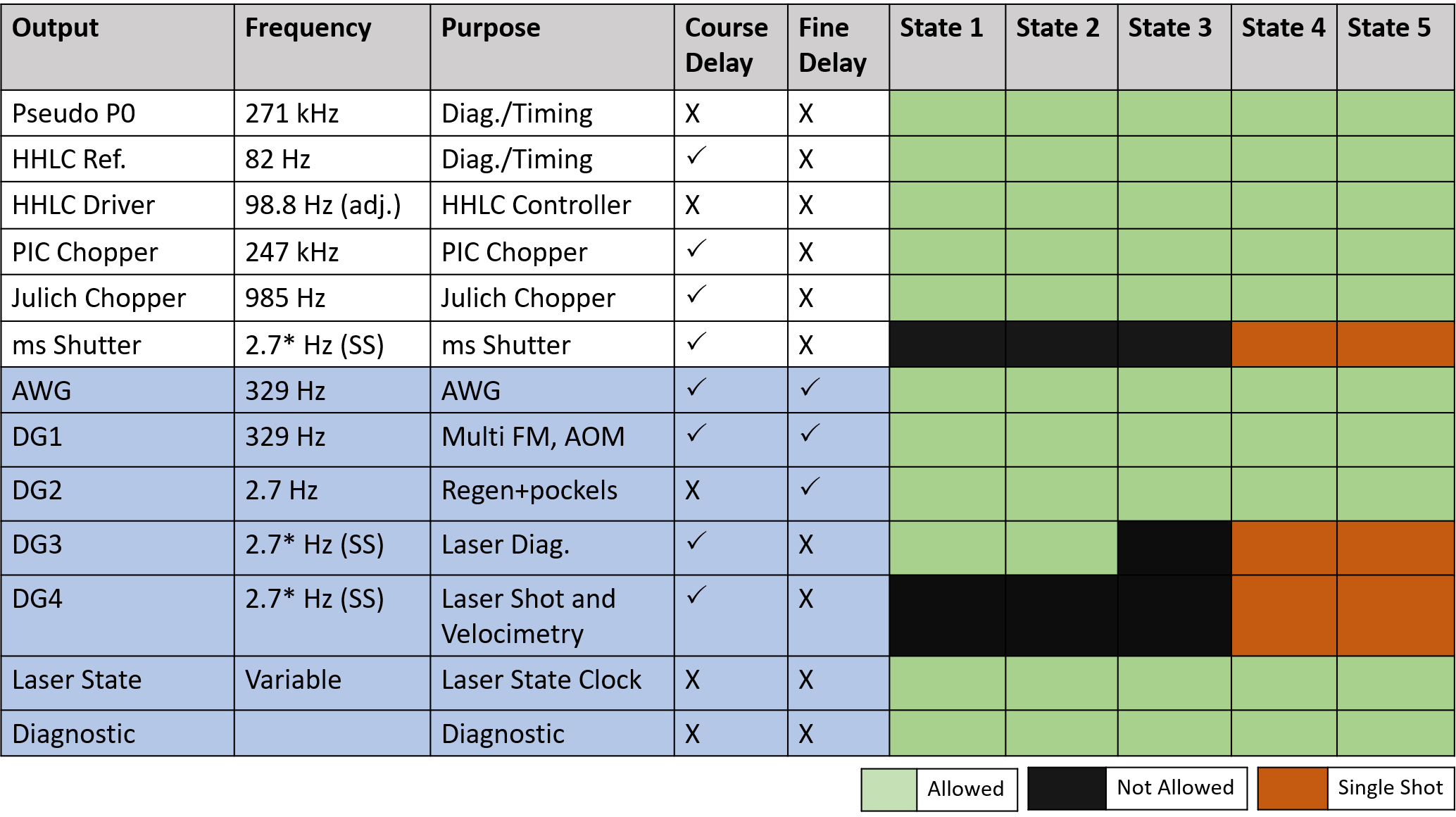}
\caption{\label{fig:Trigger_Req} Signals and triggers for the DCS timing and control system, their frequencies, delay settings and requirements.}
\end{figure*}

The triggers to fire the final amplification stages of the laser, as well as the triggers for single-shot diagnostics (e.g. VISAR), cannot be fired at 2.7 Hz, but must be derived from the 2.7 Hz clock controlling the previous laser stages to have adequate jitter with respect to prior stages. Consequently, these triggers are gated 2.7 Hz signals,  and the gate logic is determined by a firing state system, explained in the next section. All triggers are separately controllable by a combination of the FPGA and a suite of delay generators (Stanford Research Systems, DG645) with r.m.s. jitter <25 \emph{ps}. The phase of the internal FPGA clock relative to the master RF clock is controlled with a precision of 17 \emph{ps}. The various triggers, their purpose, frequency, and state gating logic are shown in Fig.~\ref{fig:Trigger_Req}.

\subsection{\label{sec2b}FPGA Firmware}

\begin{figure*}
\includegraphics[scale=0.6]{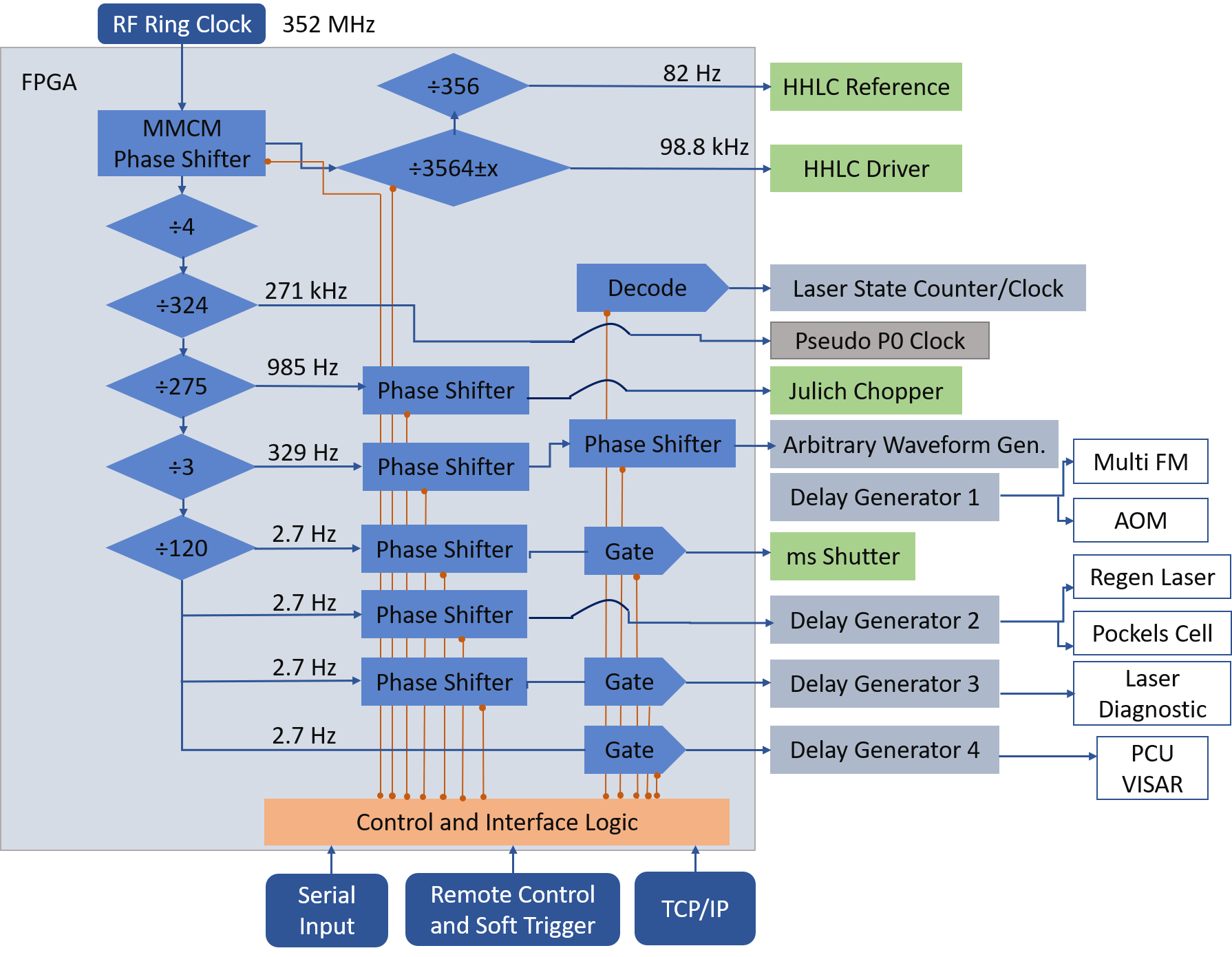}
\caption{\label{fig:TTC_BD}A block diagram representing the blocks of timing and control logic.}
\end{figure*}

The triggering consists of several states which are set through the laser control system software and the delays are set using a user-friendly FPGA controller software. The delay/phase shifting is controlled through the network via ethernet. The timing and control system block diagram is shown in Fig.~\ref{fig:TTC_BD}. Vivado 2018.2 is used for the firmware implementation. The FPGA synchronization was derived from the previous work at BioCARS [\onlinecite{BioCARS1, BioCARS2}]. The Zynq FPGA SoC (system on chip) architecture is divided into two parts: the processing system (PS) and the programmable logic (PL). The PS is the main processing unit and includes an ARM Cortex-A9 processor, on-chip memory, and various peripherals whereas the PL consists of a number of hardware accelerators. Between the PS and PL, an interconnection is used to allow the system to communicate. Advanced eXtensible Interface (AXI) protocol has been used to connect various Intellectual Property (IP) cores. AXI4-Lite protocol is used for simple, low-throughput memory-mapped communication between control and status registers. 

The various blocks of firmware and their functionality are described below:

\textbf{APS RF Clock Input}: The 352 MHz ring clock comes from a bunch clock generator LVDS output from a VME crate [\onlinecite{BunchClock}]. This is fed into the FPGA through the clock pins on a XM105 Debug card. Once in the FPGA, the ring clock goes first into the MMCM primitive. This primitive synchronizes the FPGA logic and allows 17 \emph{ps} resolution shifting of the FPGA clock tree with respect to the ring clock using the “Fine shift” feature. 

\textbf{Clock Divider}: The 352 MHz FPGA clock then gets divided down into the desired frequencies to generate various triggers. Custom frequency divider logic generates the desired frequencies. It generates an output pulse at a lower frequency than the master clock to drive a clock enable input of a register or a counter. The module uses parameters to define the frequency division divisor and the width of the internal clock pulse counter. 

\textbf{Phase Shifter or Delay}: The output clocks from the clock divider go into the inputs of the “Phase Shifters” or “delays”.  Most “delays” are several adjustable length shift registers cascaded together and clocked at different rates using the clocks that were generated in programmable logic. The code utilizes an included IP “c\_shift\_ram”. The delay comes from adjusting the registers to be longer/shorter, therefore it takes longer/shorter for the trigger to propagate through.

\textbf{Laser States}: After the phase shifting, the clock/trigger is gated/controlled based upon the laser state set by the laser control software. There are five laser states: 1. Idle, 2. Charging, 3. At voltage, 4. Fire and 5. Diagnostics. The laser state input is provided using serial communication and set the registers value. The trigger then exits the FPGA and either goes into an external delay generator or a Pulse Research Lab box. A laser state output clock is also implemented which sends variable frequency output clock based on the input state. This output is further sent to different devices for synchronization. 

\textbf{HHLC Logic}: The FPGA generates the 98.8 kHz HHLC clock from the 352 MHz auxiliary clock by dividing by 3564. If the phase shift of the HHLC needs to be increased, the phase of the chopper is advanced by dividing by 3563 or slipped by dividing by 3565. It also outputs the HHLC clock with a pulse length of 1 µs. 

\begin{figure}[h!]
\begin{center}
\includegraphics[scale=0.36]{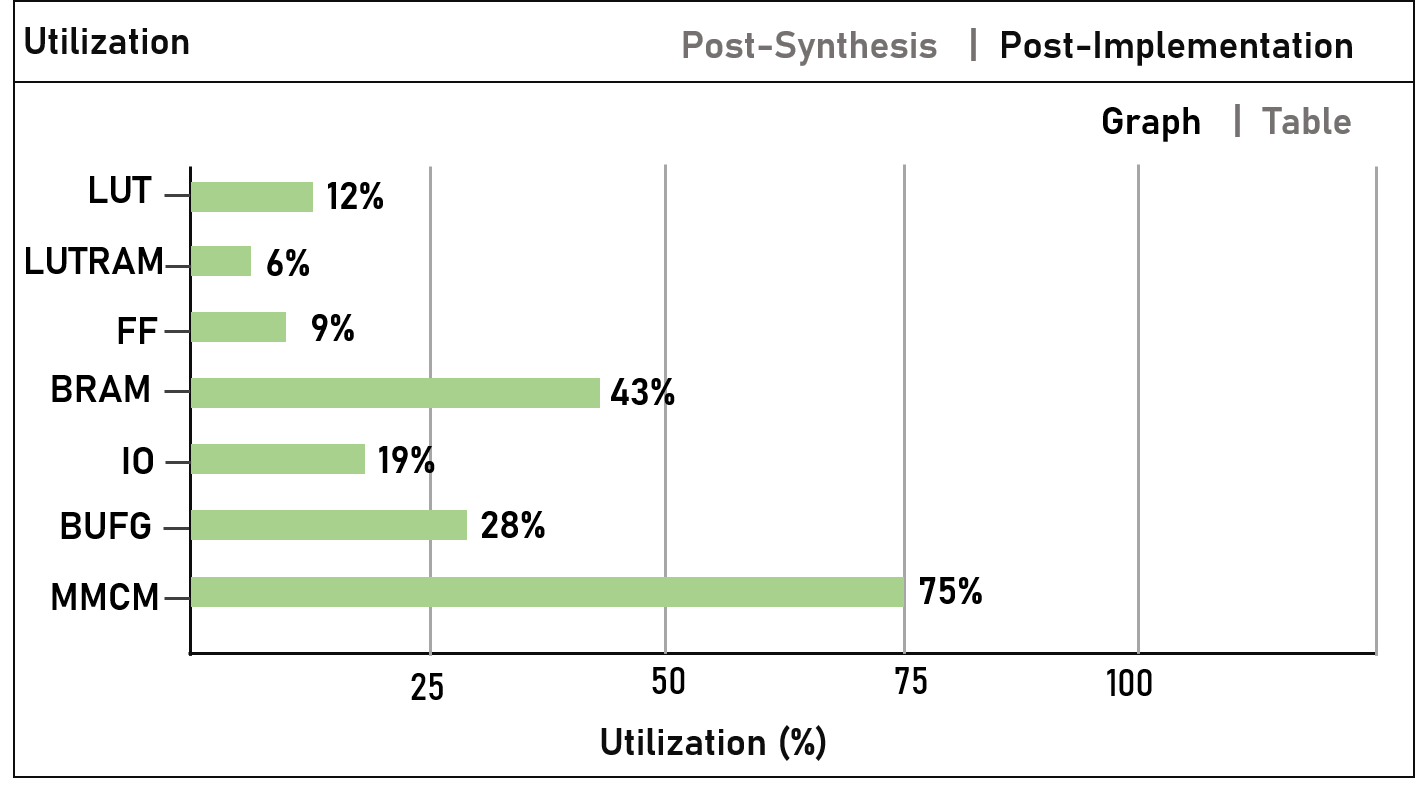}
        \caption{Device Utilization for Zynq FPGA}
        \label{fig:FPGA_Util}
\end{center}
\end{figure}

\textbf{Jülich Chopper}: Triggers for the Jülich chopper are generated using the clock divider IP and variable phase shifter blocks.
The device utilization is shown in Fig.~\ref{fig:FPGA_Util}.

\subsection{\label{sec2c} Software}

\begin{figure}[h!]
\begin{center}
\includegraphics[scale=0.36]{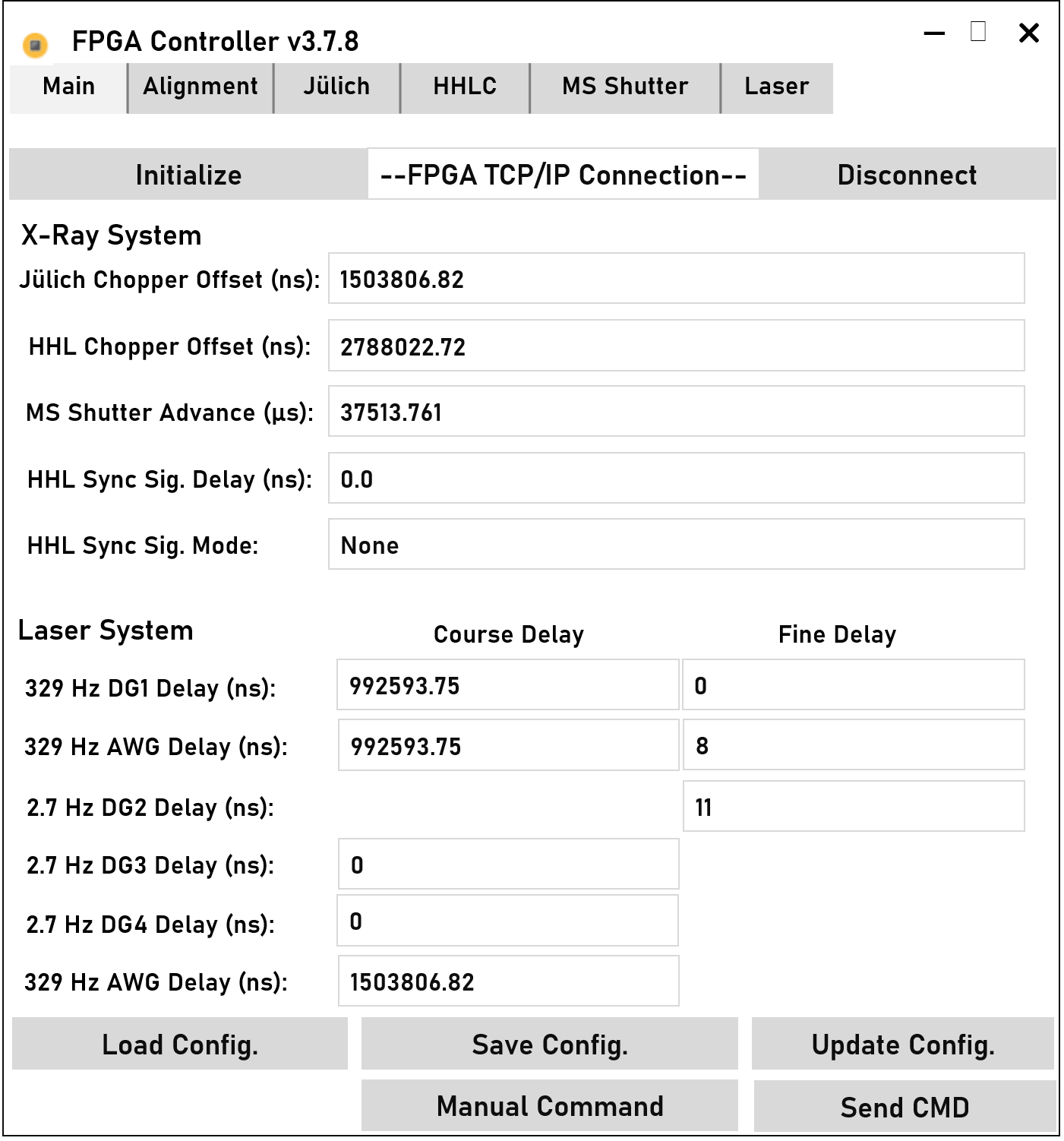}
        \caption{The FPGA Controller software GUI for the x-ray system and laser system for delay and offset settings.}
        \label{fig:Software}
\end{center}
\end{figure}

Xilinx Software Development Kit (XSDK) is used for creating embedded application-based software. The hardware specifications are imported to SDK from Xilinx Vivado to create a Hardware Platform Specification project. Lightweight IP (lwIP) is used for TCP/IP networking stack for the system. The Graphical User Interface (GUI) has been developed using PyQt which is a set of Python bindings for the Qt cross-platform C++ framework. The FPGA Controller GUI provides information about the FPGA TCP/IP connection. Also, the delays and offset for the Jülich chopper, HHLC, and MS shutter can be controlled using the FPGA Controller GUI. The GUI also provides course and fine delay settings for laser triggers with 329 Hz and 2.7 Hz clocks. Users can also set the laser state using the FPGA controller software. The screenshot of the FPGA controller GUI is presented in Fig.~\ref{fig:Software}.

\section{\label{sec3}System Details and Architecture}

\begin{figure}[h!]
\begin{center}
\includegraphics[scale=0.98]{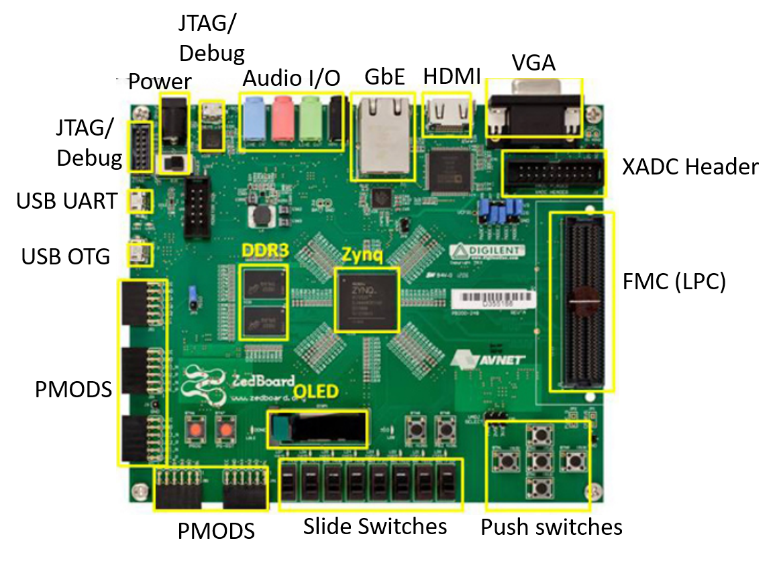}
        \caption{The Xilinx Zynq FPGA and its components.}
        \label{fig:Zynq}
\end{center}
\end{figure}

\begin{figure}[h!]
\begin{center}
\includegraphics[scale=0.32]{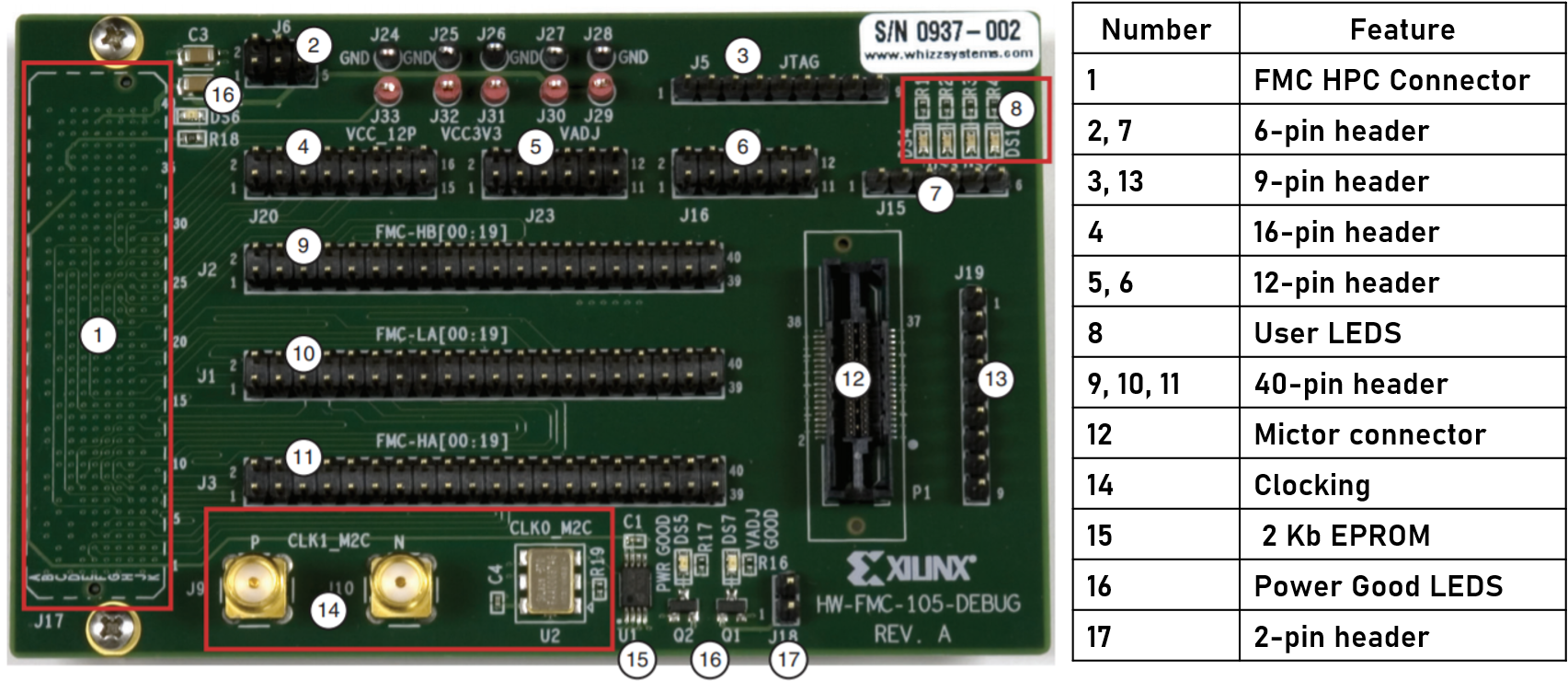}
        \caption{The Xilinx XMC Debug card and its features.}
        \label{fig:XMC}
\end{center}
\end{figure}

The timing and control digital logic has been implemented using a Xilinx Zedboard FPGA, which features a Zynq®-7000 All Programmable SoC XC7Z020-CLG484-1 FPGA board. It consists of several expansion connectors: FMC connectors, Pmod (peripheral module interface) connectors, an ethernet port, and usb connectors, which provide easy access to programmable logic inputs and outputs. The 352 MHz clock is fed into the FPGA using the Xilinx XMC Debug card. The Zedboard block diagram and FMC XM105 Debug Card are shown in Fig.~\ref{fig:Zynq} and Fig.~\ref{fig:XMC}, respectively. The XM105 is connected to the ZedBoard using the FMC connector. The various trigger signals are routed to Pmod connectors to provide signal outputs. A Pulse Research Lab 4 Channel High Input Impedance 50 Ohm TTL Line Driver is used to convert High Impedance TTL/CMOS Outputs to 50 ohm TTL Outputs. 

\section{\label{sec4}Results and Performance Evaluation}

\begin{figure*}
\includegraphics[scale=1.2]{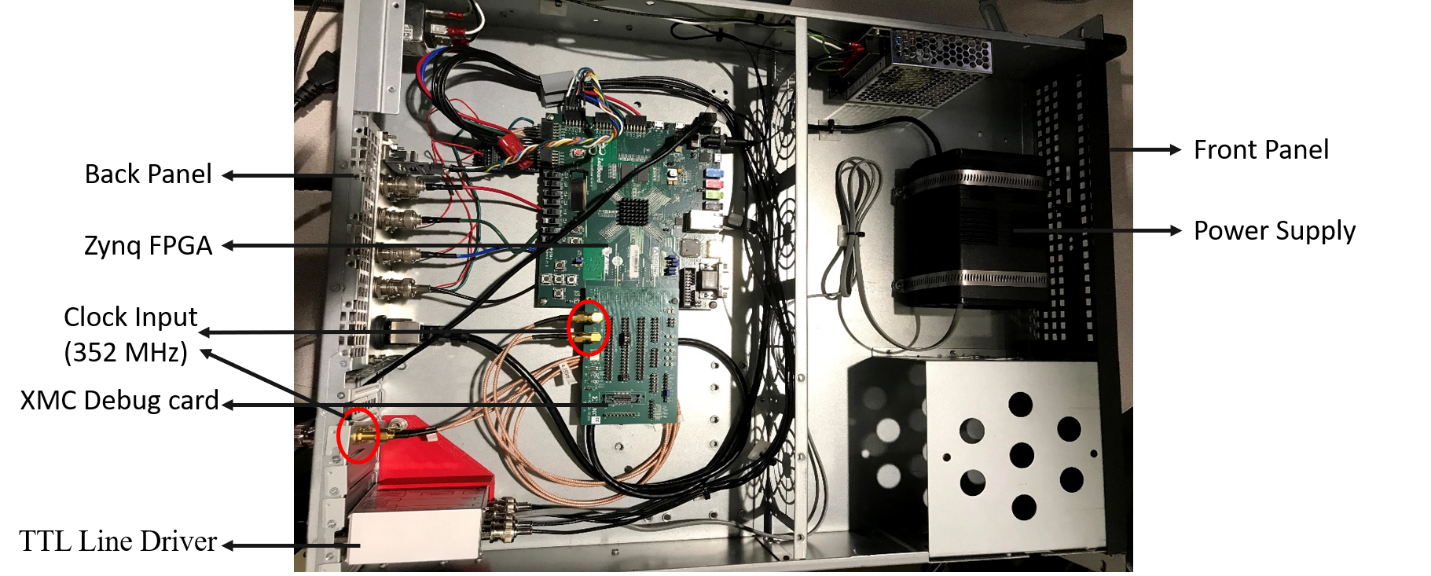}
\caption{\label{fig:DCS_System}The FPGA based timing and control system setup at DCS.}
\end{figure*}

The developed FPGA-based timing and control system is a stand-alone configurable, modular, hardware / software platform. Fig.~\ref{fig:DCS_System} shows an image of the actual system with all the components: Zynq FPGA, debug card, TTL line driver and connectors. The system is rack mounted, and the trigger outputs (BNC connectors) are routed to the various devices using coaxial cables. The performance evaluation has been conducted by measuring the relative jitter between the output trigger signal and 352 MHz input clock signal. 

The 352 MHz input is provided by the APS bunch clock generator. The bunch clock generator is a single width VME card, which provides precise timing pulses for individual bunches of interest in the stored beam structure. These timing pulses are referenced to the storage ring revolution clock signals (P0) and the 352 MHz RF signal from the low-level RF system. The laser states are set by commands sent through a serial port that control the trigger outputs. The delay/phase shifting of the triggers are set using the controller GUI and are controlled through the network via ethernet. The performance of the developed was evaluated by measuring the jitter with respect to the 352 MHz APS RF clock. For example, the jitter of the 329 Hz AWG trigger was measured using a Tektronix DPO 7254C 2.5 GHz, 40 GS/s oscilloscope, by triggering on the 329 Hz signal and measuring a histogram of the 352 MHz signal delay to reach 50\% of it’s peak voltage. The measured jitter of the 329 Hz AWG signal with respect to the 352 MHz APS RF clock is 15.5 \emph{ps} r.m.s. Similarly, the jitter was measured for the delay generator triggers and the time resolution and jitter values are reported in Table ~\ref{tab:table2}. Critically, the measured jitter of the AWG trigger easily meets the requirement of having the laser to x-ray delay jitter much lower than the x-ray probe pulse duration. 
 


\begin{table}
\caption{\label{tab:table2}Measured timing resolution and timing jitter for various signals}
\begin{ruledtabular}
\begin{tabular}{lccc}
\textbf{Signal} & \textbf{Frequency} & \textbf{Delay Resolution}  & \textbf{Timing Jitter}\\
\textbf & & \textbf& \textbf{(r.m.s.)}\\
\hline
Pseudo P0 &	271 kHz	& 11.3 \emph{ns} & - \\
HHLC Ref. &	82 Hz &	11.3 \emph{ns} & - \\
Julich Chopper &	985 Hz &	11.3 \emph{ns} & - \\
MS Shutter &	2.7 Hz &	3.7 \emph{µs} & - \\
AWG	 &329 Hz &	78 \emph{ps} &	15.5 \emph{ps} \\
DG1 &	329 Hz & 78 \emph{ps} &	16 \emph{ps} \\
DG2, DG3, DG4	& 2.7 Hz &	78 \emph{ps} &	17 \emph{ps} \\
\end{tabular}
\end{ruledtabular}
\end{table}

\section{\label{sec5}Conclusion}

An FPGA-based timing control system has been developed for the Dynamic Compression Sector at the Advanced Photon Source. The system is synchronized with the APS 352 MHz electron bunch clock and provides a precise, low-jitter trigger signals for the x-ray shutters, laser subsystems, and shot diagnostics. The delay (phase shifting) can be controlled using a custom FPGA controller software with high precision. The measured timing jitter of the AWG trigger, which determines the laser arrival time on target, is 15.5 \emph{ps} r.m.s with respect to the 352 MHz input clock, allowing the laser pulse to be timed with respect to the x-ray probe pulse with an accuracy much less than the x-ray probe pulse duration. Potential future work includes using ultrascale FPGAs to further reduce the timing uncertainties and to provide finer delay precision.

\nocite{*}
\bibliography{aipsamp}
\end{document}